E126 -- *De novo genere oscillationum*
On a new class of oscillations

L. Euler

Translated and Annotated

by

Sylvio R. Bistafa[a]

January 2021

## Foreword

In this publication, Euler derived for the first time, the differential equation of the (undamped) simple harmonic oscillator under harmonic excitation, namely, the motion of an object subjected to two acting forces, one proportional to the distance travelled, the other one varying sinusoidally with time. He then developed a general solution, making extensive use of direct and inverse sine and cosine functions. After much manipulation of the resulting equations, he proceeds to analyze the periodicity of the solutions by varying the values of the parameters, to finally finding out the phenomenon of resonance by saying ”…. Among all these cases, the one which deserves particular attention is that for which $2b = a$, in which the oscillation distance eventually grows up to infinite: this effect is most remarkable, since it is generated by finite forces.”

§1.

Although the doctrine of oscillations and of bodies in alternating motion has been subjected to much scrutiny, such that nothing new is seen possible to be discovered; nonetheless, a new kind of oscillation is put forward in this dissertation, which nobody has so far touched upon, and that still needs a particular analysis. Indeed, at first, it caught my attention to be considered in a dissertation by the most renowned colleague Krafft, described as a certain type of unusual oscillations observed in a portable suspended clock; and indeed afterwards, observed in the sea tide, by recognizing that the alternating motion of the sea belongs to the same type of oscillation.

§2. It is said that the oscillations of a body is to execute an alternating motion, when its whole or its parts are set into perpetual motion in a given space, such that they alternately advance and retard in opposite directions. Thus, indeed, there are grounds for comparison with the motion of pendulums, whose theory is seen to be equally considered in very common cases, to which it is conveniently applied to other type of oscillations, such as: vibrations of cords, trembling of bells, undulations of the waters; and also, the flux and reflux of the sea. In these, all the motions are seen to display such reciprocity and an alternate change along opposite directions.

[a] sbistafa@usp.br

§3. Therefore, since these properties are common to all oscillatory motions, whereby I will now, indeed, expose this matter to a new kind of examination that diverges from others already pursued. Then, be set a curved or straight line $ACB$ representing the space, in which the body or a portion of a body is put into reciprocal motion, with alternate changes, at one instant displaced to the right towards $ACB$, and at the other instant towards $BCA$ (Figs. 6 &7). Since there is no body that by itself once released can freely produce that kind of reciprocal motion, only to uniformly advance forward in a straight line, there is a need of forces to produce the oscillatory motion, which is the characteristic that marks the principal difference of this oscillatory motion.

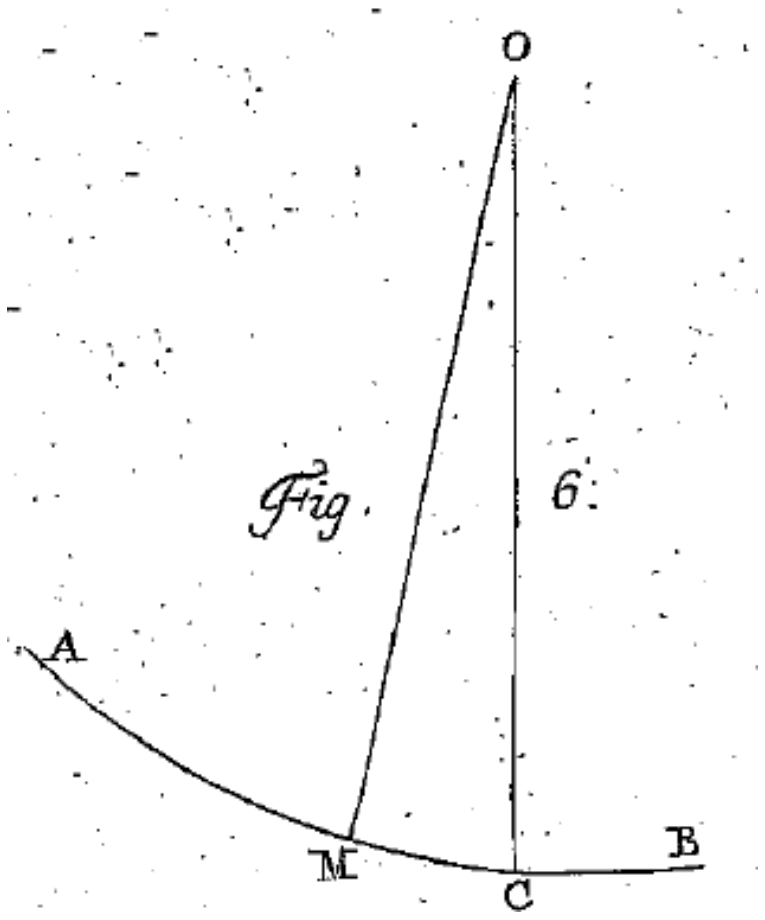

§4. However, when we look at the forces jointly with the distance in the figure in which the motion takes place, we very conveniently recognize that this distance can be depicted by the straight line $ACB$ (Fig. 7). Therefore, since the motion consists of alternations towards the right and the left, likewise should behave the forces, which at one instant drive the body towards the left and in the next instant towards the right. Hence, these forces should have extreme variations, and immediately after becoming negative; the force, indeed, can be considered as impelling towards the left, likewise as a negative force urging towards the right.

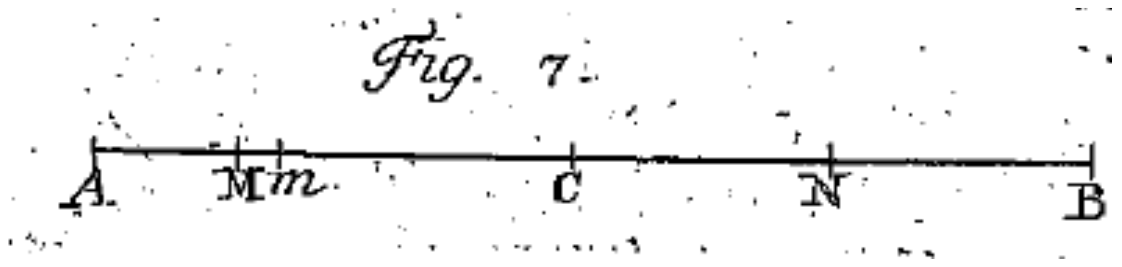

Therefore, if $p$ will be the force that drives the body while it moves around $M$, it is necessary that $p$ be a variable quantity, which, not only due the variable circumstances, should be made greater or minor, but also becoming itself negative.

§5. But if the magnitude of this force $p$ is only determined by the position that the body occupies throughout the distance $ACB$, then I call the resulting oscillations of the first kind: and this category contains all type of oscillations considered so far, which, indeed, takes place in a vacuum space. For this kind of oscillation, the force $p$ will express some function of a quantity, which depends on the location of the body $M$, certainly some function of the distance $MC$, considering that $C$ is a fixed point along the distance $ACB$. But, since such oscillations are observed as being isochronous, the force $p$ is directly proportional to the distance $MC$; which, if the body moves between $A$ and $C$, tending to the right,

whereas, for the body between $C$ and $B$, say at $N$, the force would have been negative, urging the body towards the left, that is from $N$ to $C$.

§6. I call oscillations of the second kind those that depend partly on the force $p$ by means of the distance $MC$, and partly on the velocity which the body has at $M$, in these cases, $p$ will be some function of the distance $MC$ and further, on the velocity at $M$. Belong to this category, especially those oscillations that occur in a resistant medium: this resistance is, indeed, a function proportional to the velocity, besides, the body is hindered by an absolute retarding force, considered as a resistance, being its direction, always contrary to the motion of the body. On the other hand, the type of data that must be acquired for seeking the absolute force of the law of resistance, for the oscillations to turn out isochronous, I had exposed it in more detail in my Treatise on the motion of bodies.

§7. Finally, in the third kind of oscillation as I call it, the body is driven by an absolute force that depends on the distance $MC$, besides being a force, whose magnitude is determined by a fixed time interval, whistle the body is at $M$. To my knowledge, so far, nobody has delt with this kind of oscillations; and being observed daily in the world, they are of no small importance. Indeed, to this class, belong the oscillations mentioned above, firstly observed by the most renowned Professor Krafft, in which, for the reasons that the forces producing the internal motion of the clock are time-dependent, the absolute force is originated by the weight of the clock, being proportional to distance from the point of equilibrium.

§8. Moreover, certainly, this kind of oscillations encompasses the reciprocal motion of the sea or its alternate elevation and depression. Indeed, the principal characteristic of the force promoting the sea rousing motion depends on the position of the moon, which is raised and depressed in alternations at each interval of about twelve hours: whence, this force depends neither on the position of the water nor on its velocity, but rather on the instant of time. Besides, this force is truly urged by the sea due to its own gravity, such that, it will go down, if its surface has been raised, and on the contrary, it is raised, if the surface dwells below. Wherefore, if the motion of the sea should be defined from the effect of these two forces, it will be required to investigate this third kind of oscillation from the natural causes.

§9. Then, let us consider that the oscillations occur along the distance $ACB$ (Figs. 7 & 8), and that the body while it dwells in $M$, is incited by two forces, one of which depends on the location $M$, which is proportional to the distance $MC$: therefore, this force, meanwhile the body is along $AC$, will be urged towards the right, and on the contrary, towards the left, if the body is located along $BC$. Moreover, the other force depends on time, which at some instant, the body is pushed towards the right, and at another instant towards the left, and that without any further consideration of the type of body, depends on its location. Let us consider that the time flows uniformly, moving around along the circumference $FDHE$, with the time conveniently marked along it. Further, the forces are proportional to the sines of the arcs marking the time, and they are considered positive, if they drive towards the right, and on the contrary, they are negative, towards the left.

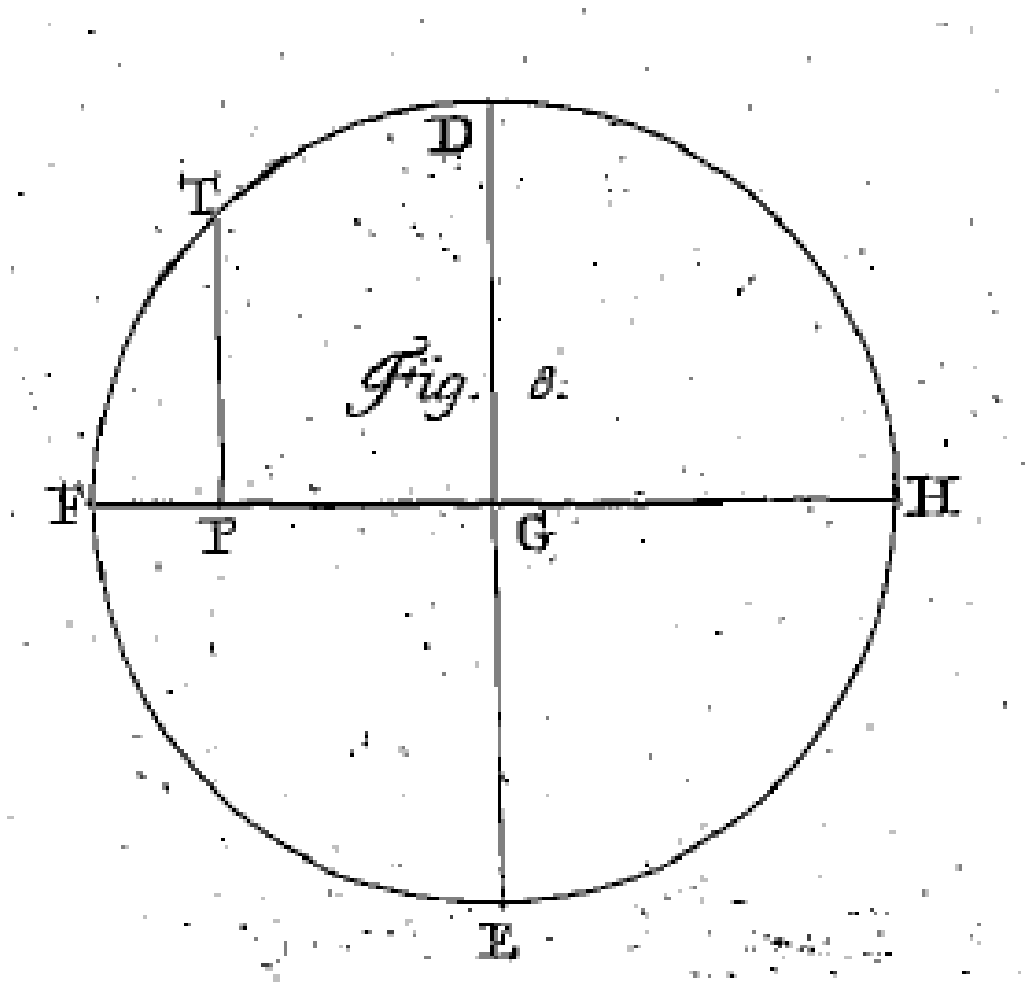

§10. Set the initial time as $F$, at the beginning of the oscillations, and with the time flowing along $FTDHE$. Then, at the beginning, the force driving the body will be null, and, after the time $FT$, the body is pushed towards the right by the force $PT$; which will be maximum after the time $FD$ has elapsed, after which, it decreases again, to complete vanish once the time $FDH$ has elapsed. Next, during the passage of time between $H$ and $F$, passing through $E$, that force will be negative, and it will attract the body towards the left; and, after the time given by all the periphery has elapsed, the same driving force will return back to the initial condition, which is necessary to generate the proposed oscillatory motion, and if these are the only acting forces: then, a priori, it will be found that an absolute force that depends on the location of the body, the oscillatory motion will be more agitated, the greater the difference in time between these forces.

§11. Be set $FG = DG = a$, as the radius of the circumference $FDHE$, and the whole circumference $FDHE = 4c$, such that $c$ denotes the forth part of the circumference: consider now that the elapsed time is represented by the arc $FT$, which, once the arc $FT$ is set as $= t$, then, [the time][b] will be $= \frac{t}{\sqrt{a}}$: and, indeed, for the sake of homogeneity, it is convenient to express the time as a function of half of the dimension of lines. Moreover, this instant of time, is the one when the body appears at the position $M$, and the distance $MC = s$; yet, at this location, the velocity is towards the left, with a value given by the height $v$. Then, at this location, a priori, it had been driven with a force towards the right, and this force, since it is proportional to the distance $MC$, is set $= \frac{s}{b}$, being the force of gravity $= 1$.

§12. The other force, depends on time and will also urge towards the right, being proportional to the sine $PT$, and accordingly, the sine of the arc $FT$ is positive. Let us set the arc $FT = t$, and the sine $PT = y$, and the force impelling the body towards the right is $= \frac{y}{g}$. Since the body at $M$ is jointly driven by these forces in the same direction towards the right, the total force will be $= \frac{s}{b} + \frac{y}{g}$; therefore, the

[b] The speed of the body in its motion is $\frac{dx}{dt}$, and if we suppose that this speed is equal [to the speed] a heavy body acquires when it falls from the height [$x$], it is necessary to take [$\left(\frac{dx}{dt}\right)^2 = x$] or [$dt = \frac{dx}{\sqrt{x}}$]; from the latter, the relation between the time $t$ and the distance $x$ is known [after integration].

acceleration that accomplishes the elemental distance $Mm$, will be known. Since, indeed, $Mm = -ds$, by the law of acceleration $dv = -ds\left(\frac{s}{b} + \frac{y}{g}\right)$, whose integration should be made since the beginning of the motion, certainly with the velocity that the body has at this location.

§13. Besides, indeed, the equation of the circumference gives $dt = \frac{ady}{\sqrt{a^2-y^2}}$, from which we have that $t = a\sin^{-1}\frac{y}{a}$, denoting $\sin^{-1}\frac{y}{a}$ the arc whose sine is $\frac{y}{a}$, having the value $= 1$ in the circumference semi-diameter: similarly, $y = a\sin\frac{t}{a}$, denoting $\sin\frac{t}{a}$, the sine of the arc $\frac{t}{a}$, whose value $= 1$ in the circumference radius. Then, if $t = c$, then, $y = a$, and if $t = 2c$, then $y = 0$; and by denoting $i$ any integer number, we have that if $t = 2ic$, then $y = 0$; when $\sin t = (4i + 1)c$, then $y = a$; whereas when $\sin t = (4i - 1)c$, then $y = -a$. Therefore, $t$ and $y$ can be freely interchanged in the computations.

§14. Wherefore, we have four unknowns to be found, and three equations; two of which have already been disclosed. In fact, the third equation will be generated from the introduction of time. Since, indeed, the total time $= \frac{t}{\sqrt{a}}$, the infinitesimal time that it takes to complete the infinitesimal distance $Mm$ is $= \frac{dt}{\sqrt{a}}$, and, indeed, the same time is also $= -\frac{ds}{\sqrt{v}}$, whence giving this equation $\sqrt{v} = \frac{-ds\sqrt{a}}{dt}$. Therefore, we have four variables , $s, t, y, and\ v$, and three equations $\boldsymbol{dv = -ds\left(\frac{s}{b} + \frac{y}{g}\right)}$; $\boldsymbol{dt = \frac{ady}{\sqrt{a^2-y^2}}}$ and $\boldsymbol{\sqrt{v} = \frac{-ds\sqrt{a}}{dt}}$,[c] any two of these possible to be combined into one equation.

§15. However, the oscillation will be conveniently known, if it will be possible to locate the body along $AB$ at any given instant of time. Then, the variables $y$ and $v$ can be conveniently eliminated, resulting in one equation between $s$ and $t$. Since, indeed, from two of the previous equations $y = a\sin^{-1}\frac{t}{a}$ and $v = \frac{-ads^2}{dt^2}$, which, after substituting into the equation $dv = -ds\left(\frac{s}{b} + \frac{y}{g}\right)$, results in the following equation between $s$ and $t$: $\frac{2adsdds}{dt^2} = -ds\left(\frac{s}{b} + \frac{a}{g}\sin\frac{t}{a}\right)$ or $\boldsymbol{2adds + \frac{sdt^2}{b} + \frac{adt^2}{g}\sin\frac{t}{a} = 0}$[d] : which, in fact, can be integrated twice, resulting in a particular equation between $s$ and $t$.

§16. Before beginning the integration of this equation, which certainly is not a small difficulty, it will be useful to assess carefully the non-zero special cases. And, in fact, the first to vanish, is the internal force that depends on the instant of time, such that the body is only excited by the force $\frac{s}{b}$, which depends on the distance $MC$. By assuming $g = \infty$, then we will have the equation $2abdds + sdt^2 = 0$, which, once multiplied by $ds$ and integrated gives $abds^2 + \frac{s^2dt^2}{2} = \frac{Cdt^2}{2}$; or $\frac{-ds\sqrt{a}}{dt} = \sqrt{\frac{C-s^2}{2b}} = \sqrt{v}$, in which the value

---

[c] Key expressions are highlighted in bold.

[d] Here it is in a slightly modernized form $2a\frac{d^2s}{dt^2} + \frac{1}{b}s + \frac{a}{g}\sin\frac{t}{a} = 0$. The equation for the one-dimensional motion of a body of mass $m$, subjected to an ideal spring of strength $k$ and to friction of strength $c$, and acted upon externally by a driving force of amplitude $F$, that varies harmonically (i.e. sinusoidally) with frequency $\omega$, in time $t$, reads: $m\frac{d^2s}{dt^2} + ks + c\frac{ds}{dt} = F\sin\omega t$. It can be seen that Euler had develop the equation for the harmonic mechanical oscillator without friction.

of the constant $C$ is determined such that the initial velocity corresponds to as follows. Let us consider that the velocity at $C$ is due to the height $b$, then, $C = 2b^2$, resulting in the following equation $\frac{-ds\sqrt{2ab}}{\sqrt{(2b^2-s^2)}} = dt$.

§17. The integration of the last equation can be approached by the quadrature of the circle, resulting in $t = C - \sqrt{2ab} \cdot \sin^{-1} \frac{s}{\sqrt{2b^2}}$. However, since $\sqrt{v} = \sqrt{\frac{2b^2-s^2}{2b}}$; the beginning of the motion, in which the velocity had vanished, falls upon the point $A$, giving $CA = \sqrt{2b^2}$ : from which the constant is given by $C = \sqrt{2ab} \sin^{-1} 1$. Wherefore, during this time the distance $AM$ or $\boldsymbol{t = \sqrt{2ab} \cdot \cos^{-1} \frac{s}{\sqrt{2b^2}}}$ : and, since time $= \frac{t}{\sqrt{a}}$ , the equation for the time itself is $\frac{t}{\sqrt{a}} = \sqrt{2b} \cdot \cos^{-1} \frac{s}{\sqrt{2b^2}}$. The time that it takes for the body to reach the medium point $C$ from $A$ is $= \sqrt{2b} \cdot \cos^{-1} 0 = \frac{\pi\sqrt{2b}}{2} = \frac{\pi\sqrt{b}}{\sqrt{2}}$, denoting $\pi$ the circumference whose diameter $= 1$. Whence, it is seen not only the nature of these oscillations but also their isochronism.

§18. Now, let us consider to vanish the force which is proportional to the distance $MC$, and that the body is only urged by the other [force] that is time-dependent; this condition being satisfied by putting $b = \infty$, resulting in the following equation $2gdds + dt^2 \sin\frac{t}{a} = 0$. For the integration, it will be helpful to note that the $differential\ of$ $\sin\frac{t}{a} = \frac{dt}{a}\cos\frac{t}{a}$ and that the $differential\ of$ $\cos\frac{t}{a} = -\frac{dt}{a}\sin\frac{t}{a}$. Then, considering the second identity, the first integration will give $2gds - adt\cos\frac{t}{a} = Cdt$, whence $\frac{ds}{dt} = \frac{C + a\cos\frac{t}{a}}{2g}$, and the velocity in $M = \sqrt{v} = \frac{-ds\sqrt{a}}{dt} = \frac{-Ca - a^2\cos\frac{t}{a}}{2g\sqrt{a}}$. By considering that at the initial time the velocity is towards the right, and due to the height $b$, then $2g\sqrt{ab} = -Ca - a^2$, and, therefore, $Ca = -a^2 - 2g\sqrt{ab}$, from which, the velocity after the time $\frac{t}{\sqrt{a}}$ at the same location will be $\sqrt{v} = \sqrt{b} + \frac{a\sqrt{a}\left(1-\cos\frac{t}{a}\right)}{2g}$.[e]

§19. Whenever the sine of the arc is positive, it is understood that the velocity is always positive or directed towards the right, if, indeed, at the initial time the velocity at the same location is $\sqrt{b}$. Then, in this case, the body will advance without bounds along $AB$, in fact, in a non-uniform motion; truly, at the elapsed times $\frac{0}{\sqrt{a}}$; $\frac{4c}{\sqrt{a}}$; $\frac{8c}{\sqrt{a}}$; $\frac{12c}{\sqrt{a}}$ and generically $\frac{4ic}{\sqrt{a}}$, the velocity will be $= \sqrt{b}$; however, for the elapsed times $\frac{c}{\sqrt{a}}$; $\frac{3c}{\sqrt{a}}$; $\frac{5c}{\sqrt{a}}$ and generically $\frac{(2i+1)c}{\sqrt{a}}$, the velocity will be $= \sqrt{b} + \frac{a\sqrt{a}}{2g}$; finally, for the times $\frac{2c}{\sqrt{a}}$; $\frac{6c}{\sqrt{a}}$; $\frac{10c}{\sqrt{a}}$ and generically $\frac{2(2i+1)c}{\sqrt{a}}$, the velocity will be maximum and $= \sqrt{b} + \frac{a\sqrt{a}}{g}$. Wherefore, if the initial velocity $\sqrt{b}$ is not negative or stretching out towards the left, the motion will not be reciprocal, and there will be no oscillations.

§20. Therefore, for the body to acquire a perpetual oscillatory motion in the same interval, in which it alternately advances and retards, it is necessary that the velocity be equally negative and positive: which

---

[e] This is not the same expression that appears in the manuscript, and seems to be the correct result of the integration, which was later on confirmed in §20.

will come forth if the body initially moves towards the right with a velocity $=\frac{a\sqrt{a}}{2g}$: or by putting $\sqrt{b}=\frac{-a\sqrt{a}}{2g}$. Indeed, under this hypothesis, at a given time $\frac{t}{\sqrt{a}}$, the velocity will be towards the left, and given by $\sqrt{v}=\frac{-a\sqrt{a}\cos\frac{t}{a}}{2g}$. Therefore, for the times $\frac{0}{\sqrt{a}}$; $\frac{4c}{\sqrt{a}}$; $\frac{8c}{\sqrt{a}}$ and generically $\frac{4ic}{\sqrt{a}}$, the velocity will be $=\frac{-a\sqrt{a}}{2g}$; for the times $\frac{c}{\sqrt{a}}$; $\frac{5c}{\sqrt{a}}$; $\frac{9c}{\sqrt{a}}$ and generically $\frac{(4i+1)c}{\sqrt{a}}$, and, likewise, for the times $\frac{3c}{\sqrt{a}}$; $\frac{7c}{\sqrt{a}}$ and generically $\frac{(4i+3)c}{\sqrt{a}}$, the velocity will be $=0$. Finally, for the times $\frac{2c}{\sqrt{a}}$; $\frac{6c}{\sqrt{a}}$ and generically $\frac{(4i+2)c}{\sqrt{a}}$, the velocity will be $=\frac{a\sqrt{a}}{2g}$.

§21. Thus, since the case for a regular oscillation occurs when $\sqrt{v}=\frac{-a\sqrt{a}}{2g}\cos\frac{t}{a}$; then, $\frac{ds}{dt}=\frac{a\cos\frac{t}{a}}{2g}$, or else, $2gds=adt\cos\frac{t}{a}$, which upon integration yields $2gs=C+a^2\sin\frac{t}{a}$. Be set the constant $C=0$, which is the distance $s$ that is calculated at the medium point $C$, resulting in $\boldsymbol{s=\frac{a^2}{2g}\sin\frac{t}{a}}$, which turns out repeatedly negative as well as positive. Then, for the times $\frac{0}{\sqrt{a}}$; $\frac{2c}{\sqrt{a}}$; $\frac{4c}{\sqrt{a}}$ and $\frac{2ic}{\sqrt{a}}$, the body is at point $C$. Indeed, for the times $\frac{c}{\sqrt{a}}$; $\frac{5c}{\sqrt{a}}$ and generically $\frac{(4i+1)c}{\sqrt{a}}$ the body will move around A, with $CA=\frac{a^2}{2g}$. In fact, for the times $\frac{3c}{\sqrt{a}}$; $\frac{7c}{\sqrt{a}}$ and generically $\frac{(4i+3)c}{\sqrt{a}}$, the location of the body will be at $B$, with $CB=\frac{a^2}{2g}$. Finally, the time for which the body reaches out $B$ from $A$, as well as $A$ from $B$ will be $=\frac{2c}{\sqrt{a}}=\pi\sqrt{a}$,[f] denoting $1:\pi$ the ration between the diameter and the circumference.

§22. Therefore, with this case being unfolded, it is now sufficiently realized how the integration of the differential equation $2adds+\frac{sdt^2}{b}+\frac{adt^2}{g}\sin\frac{t}{a}=0$ should be accomplished, from which, the equation of the motion should be derived, if the body is jointly driven by both forces. Indeed, as I am used to handle it, let us first accomplish the integration in the following way, in which there are no other variables beyond one dimension in a differential equation of any degree. Although, in this procedure, the construction of the equation will be done manually, nonetheless, it will result in quite integrable formulas, such that it should have priority over any other particular integration method.

§23. Indeed, my method is firstly handled as follows: reject all terms in which nowhere a variable has more than one dimension, the residual equation is then integrated[g]. Accordingly, this is the resulting equation $2adds+\frac{sdt^2}{b}=0$, which is our first case handled before, and once integrated twice results in $t=\sqrt{2ab}\cdot\cos^{-1}\frac{s}{C}$, which gives $s=C\cos\frac{t}{\sqrt{2ab}}$. The value of $s$ will be found from a rule postulated by me as follows: the value of $s$ will be brought forth in terms of new variable, such that $\boldsymbol{s=u\cos\frac{t}{\sqrt{2ab}}}$, then, $ds=du\cos\frac{t}{\sqrt{2ab}}-\frac{udt}{\sqrt{2ab}}\sin\frac{t}{\sqrt{2ab}}$; and also $dds=ddu\cos\frac{t}{\sqrt{2ab}}-\frac{2dtdu}{\sqrt{2ab}}\sin\frac{t}{\sqrt{2ab}}-\frac{udt^2}{2ab}\cos\frac{t}{\sqrt{2ab}}$.

---

[f] Since time $=\frac{t}{\sqrt{a}}$, then for the time $=\pi\sqrt{a}$, $t=\pi$, meaning that every time that an arc equals $\pi$ is completed, the body covers the distance $BA$ or $AB$.

[g] According to this rule, the term $\frac{adt^2}{g}\sin\frac{t}{a}$ should be rejected.

§24. If now those values are substituted into the proposed equation $2adds+\frac{sdt^2}{b}+\frac{adt^2}{g}\sin\frac{t}{a}=0$, the following equation will be obtained $2addu\cos\frac{t}{\sqrt{2ab}}-\frac{4adudt}{\sqrt{2ab}}\sin\frac{t}{\sqrt{2ab}}+\frac{adt^2}{g}\sin\frac{t}{a}=0$ [h]. Since the variable $u$ is not involved in the equation, be set $du=pdt$, and then, the proposed equation will be transformed into this differential of the first degree $2adp\cos\frac{t}{\sqrt{2ab}}-\frac{4apdt}{\sqrt{2ab}}\sin\frac{t}{\sqrt{2ab}}+\frac{adt}{g}\sin\frac{t}{a}=0$: which can be further transformed into $dp-\frac{2pdt}{\sqrt{2ab}}\tan\frac{t}{\sqrt{2ab}}=-\frac{dt}{2g}\cdot\frac{\sin\frac{t}{a}}{\cos\frac{t}{\sqrt{2ab}}}$, which is a more suitable form for integration.

§25. Since we have that $\frac{-dt}{\sqrt{2ab}}\sin\frac{t}{\sqrt{2ab}}=differential\ of\cos\frac{t}{\sqrt{2ab}}$, then, the last equation is transformed into $dp-\frac{2p\cdot diff.\ \cos\frac{t}{\sqrt{2ab}}}{\cos\frac{t}{\sqrt{2ab}}}=\frac{-dt}{2g}\cdot\frac{\sin\frac{t}{a}}{\cos\frac{t}{\sqrt{2ab}}}$, which becomes integrable, when multiplied by $\cos\frac{t}{\sqrt{2ab}}\cdot\cos\frac{t}{\sqrt{2ab}}$, and after the integration results in $p\cos\frac{t}{\sqrt{2ab}}\cdot\cos\frac{t}{\sqrt{2ab}}=C-\frac{1}{2g}\int dt\sin\frac{t}{a}\cos\frac{t}{\sqrt{2ab}}$, and, if the constant is enveloped into the integral, then, $p=\frac{-1}{2g\cos\frac{t}{\sqrt{2ab}}\cdot\cos\frac{t}{\sqrt{2ab}}}\int dt\sin\frac{t}{a}\cos\frac{t}{\sqrt{2ab}}$. [i] Since $p$ can be found from $t$, then $u=\int pdt$, and, finally, $s=\cos\frac{t}{\sqrt{2ab}}\cdot\int pdt$.

§26. However, not only the latter, but also the former integration is seen rather difficult, yet, a preliminary examination allows one of these integrations to be suitably. Indeed, consider the transmutation of the integral $\int dt\sin\frac{t}{a}\cos\frac{t}{\sqrt{2ab}}=\sqrt{2ab}\sin\frac{t}{a}\sin\frac{t}{\sqrt{2ab}}-\frac{\sqrt{2ab}}{a}\int dt\cos\frac{t}{a}\cdot\sin\frac{t}{\sqrt{2ab}}=\sqrt{2ab}\sin\frac{t}{a}\sin\frac{t}{\sqrt{2ab}}+2b\cos\frac{t}{a}\cdot\cos\frac{t}{\sqrt{2ab}}+\frac{2b}{a}\int dt\sin\frac{t}{a}\cdot\cos\frac{t}{\sqrt{2ab}}$ [j], and, since this last integral is similar to the one that was initially given, then, we will have that $\int dt\sin\frac{t}{a}\cdot\cos\frac{t}{\sqrt{2ab}}=\frac{a\sqrt{2ab}\cdot\sin\frac{t}{a}\cdot\sin\frac{t}{\sqrt{2ab}}+2ab\cos\frac{t}{a}\cdot\cos\frac{t}{\sqrt{2ab}}}{a-2b}+C$, whence, $p=\frac{C}{\cos\frac{t}{\sqrt{2ab}}\cdot\cos\frac{t}{\sqrt{2ab}}}-\frac{a\sqrt{2ab}\cdot\sin\frac{t}{a}\cdot\sin\frac{t}{\sqrt{2ab}}-2ab\cos\frac{t}{a}\cdot\cos\frac{t}{\sqrt{2ab}}}{2g(a-2b)\cos\frac{t}{\sqrt{2ab}}\cdot\cos\frac{t}{\sqrt{2ab}}}$, [k] which shows the value of $p$ expressed in terms of finite quantities.

§27. Since, furthermore, $u=\int pdt$, let the value of $p$ just found be multiplied by $dt$, which will make every term integrable, resulting in $u=D+\frac{C\sin\frac{t}{\sqrt{2ab}}}{\cos\frac{t}{\sqrt{2ab}}}-\frac{a^2b\sin\frac{t}{a}}{g(a-2b)\cos\frac{t}{\sqrt{2ab}}}$. Since $s=u\cos\frac{t}{\sqrt{2ab}}$, resulting, finally, in the following equation $s=D\cos\frac{t}{\sqrt{2ab}}+C\sin\frac{t}{\sqrt{2ab}}-\frac{a^2b\sin\frac{t}{a}}{g(a-2b)}$, in which the value of the

[h] The terms $-\frac{udt^2}{2ab}\cos\frac{t}{\sqrt{2ab}}$ and $\frac{sdt^2}{b}$ were rejected into this equation.

[i] This is, in fact, a slightly different result from the one that appears in the manuscript.

[j] Here, integration by parts was applied twice.

[k] $u=2a^2b\sin\frac{t}{a}, du=2ab\cos\frac{t}{a}dt$; $v=\cos\frac{t}{\sqrt{2ab}}, dv=-\frac{1}{\sqrt{2ab}}\sin\frac{t}{\sqrt{2ab}}dt$; $d\left(\frac{u}{v}\right)=\frac{udv-vdu}{v^2}\Rightarrow d\left(\frac{2a^2b\sin\frac{t}{a}}{\cos\frac{t}{\sqrt{2ab}}}\right)=$

$$\frac{-2a^2b\sin\frac{t}{a}\cdot\frac{1}{\sqrt{2ab}}\sin\frac{t}{\sqrt{2ab}}-\cos\frac{t}{\sqrt{2ab}}\cdot 2ab\cos\frac{t}{a}}{\cos\frac{t}{\sqrt{2ab}}\cos\frac{t}{\sqrt{2ab}}}dt$$

constants should be defined according to the proposed case. From this result, the velocity $\sqrt{v}$ can be easily defined, because, since it is $= \frac{-ds\sqrt{a}}{dt}$, then, $\sqrt{v} = \frac{D}{\sqrt{2b}} \sin\frac{t}{\sqrt{2ab}} - \frac{C}{\sqrt{2b}} \cos\frac{t}{\sqrt{2ab}} - \frac{ab\sqrt{a}\cos\frac{t}{a}}{g(a-2b)}$. From these equations, it will be possible to determine at any given time, the location of the body along the line $AB$ as well as its velocity.

§28. For the case where $2b = a$ or $\sqrt{2ab} = a$, the integration assumes a peculiar character, and the previous approach cannot be applied to the present case. Since we have that $\int dt \sin\frac{t}{a}\cos\frac{t}{a} = \frac{1}{2} a \sin\frac{t}{a} \cdot \sin\frac{t}{a}$, therefore, $p = \frac{C}{\cos\frac{t}{\sqrt{2ab}} \cdot \cos\frac{t}{\sqrt{2ab}}} - \frac{a\sin\frac{t}{a}\cdot\sin\frac{t}{a}}{4g\cos\frac{t}{a}\cdot\cos\frac{t}{a}}$. Whence, $\int pdt = u = \frac{C\sin\frac{t}{a}}{\cos\frac{t}{a}} - \frac{a^2\sin\frac{t}{a}}{4g\cos\frac{t}{a}} + \frac{at}{4g} + D$. Consequently, $s = D\cos\frac{t}{a} + C\sin\frac{t}{a} + \frac{at}{4g}\cos\frac{t}{a}$, with a changed constant $C$. From this result, it follows that $\sqrt{v} = \frac{-ds\sqrt{a}}{dt} = \frac{D}{\sqrt{a}}\sin\frac{t}{a} - \frac{C}{\sqrt{a}}\cos\frac{t}{a} - \frac{a\sqrt{a}}{4g}\cos\frac{t}{a} + \frac{t\sqrt{a}}{4g}\sin\frac{t}{a}$. From these, it can be seen that after an infinite time, those oscillations grow abnormally out to infinite and will extend to an infinite great distance.

§29. Since these integrations are rather out of the ordinary, and, therefore, not easily accomplished, I will present another particular method, whose power could elicit a solution to these integral equations. Since the proposed equation reads $2adds + \frac{sdt^2}{b} + \frac{adt^2}{g}\sin\frac{t}{a} = 0$, when expanded in a series of sine of the arc $\frac{t}{a}$ it will transform into $2adds + \frac{sdt^2}{b} + \frac{dt^2}{g}\left(t - \frac{t^3}{1\cdot2\cdot3a^2} + \frac{t^5}{1\cdot2\cdot3\cdot4\cdot5a^4} - \frac{t^5}{1\cdot2\cdot\ldots\cdot7a^6} + etc.\right) = 0$. Now, set $s$ as this indefinite value, $s = \alpha + \beta t + \gamma t^2 + \delta t^3 + \epsilon t^4 + \zeta t^5 + \eta t^6 + etc.$, giving the following expressions after the due substitutions are made:

$$\frac{2adds}{dt^2} = 2\cdot1\cdot2\gamma a + 2\cdot2\cdot3\delta at + 2\cdot3\cdot4\epsilon at^2 + 2\cdot4\cdot5\zeta at^3 + 2\cdot5\cdot6\eta at^4 + etc.$$

$$\frac{s}{b} = \frac{\alpha}{b} + \frac{\beta t}{b} + \frac{\gamma t^2}{b} + \frac{\delta t^3}{b} + \frac{\epsilon t^4}{b} + etc.$$

$$\frac{a}{g}\sin\frac{t}{a} = \frac{t}{g} - \frac{t^3}{1\cdot2\cdot3a^2g} + etc.$$

§30. If now the homogeneous terms of these three series are set $= 0$, the assumed coefficients of the series which was set equal to $s$, will be defined such as:

$$\gamma = \frac{-\alpha}{1\cdot2\cdot2ab}; \delta = \frac{-b-\beta g}{2\cdot3g\cdot2ab}; \ \varepsilon = \frac{\alpha}{1\cdot2\cdot3\cdot4\cdot2^2a^2b^2}$$

$$\zeta = \frac{2b + a + \frac{\beta ag}{b}}{1\cdot2\cdot3\cdot4\cdot5\cdot2^2\cdot a^3gb}; \ \eta = \frac{-\alpha}{1\cdot2\cdot3\cdot4\cdot5\cdot6\cdot2^3\cdot a^3b^3}$$

$$\theta = \frac{-b - \frac{a}{2} - \frac{a^2}{4b} - \frac{\beta a^2 g}{4b^2}}{1\cdot2\cdot3\cdot\ldots\ldots\cdot7\cdot2a^5bg}; \ \vartheta^{\mathrm{I}} = \frac{\alpha}{1\cdot2\cdot\ldots\ldots\cdot8\cdot2^4a^4b^4}$$

[I] An apparent typo in which 2 appears in place of $\vartheta$ in the original manuscript.

$$\kappa=\frac{b+\frac{a}{2}+\frac{a^2}{4b}+\frac{a^3}{8b^2}+\frac{\beta a^3 g}{8b^3}}{1\cdot 2\cdot 3\cdot\ldots\ldots\cdot 9\cdot 2a^7 bg};\ \lambda=\frac{-\alpha}{1\cdot 2\cdot\ldots\ldots\cdot 10\cdot 2^5 a^5 b^5}$$

$$\mu=\frac{-b-\frac{a}{2}-\frac{a^2}{4b}-\frac{a^3}{8b^2}-\frac{a^4}{16b^3}-\frac{\beta a^4 g}{16b^4}}{1\cdot 2\cdot 3\cdot\ldots\ldots\cdot 11\cdot 2a^9 bg};etc.$$

whence, the values of the remaining coefficients will be known.

§31. Indeed, the coefficients of even powers in $t$ progress quite regularly, whereas, the exponents of odd powers are reduced to the following forms.

$$\beta=\beta; \qquad \delta=\frac{-a+2b}{1\cdot 2\cdot 3\cdot 2ag(a-2b)}-\frac{\beta}{1\cdot 2\cdot 2ab}$$

$$\zeta=\frac{a^2-4b^2}{1\cdot 2\cdot 3\cdot 4\cdot 5\cdot 4a^3 bg(a-2b)}+\frac{\beta}{1\cdot 2\cdot 3\cdot 4\cdot 5\cdot 4a^2 b^2}$$

$$\theta=\frac{-a^3+8b^3}{1\cdot 2\cdot 3\cdot\ldots\ldots\cdot 7\cdot 8a^5 b^2 g(a-2b)}-\frac{\beta}{1\cdot 2\cdot 3\cdot\ldots\ldots\cdot 7\cdot 8a^3 b^3}$$

$$\kappa=\frac{a^4-16b^4}{1\cdot 2\cdot 3\cdot\ldots\ldots\cdot 9\cdot 16a^7 b^3 g(a-2b)}+\frac{\beta}{1\cdot 2\cdot\ldots\ldots\cdot 9\cdot 16a^4 b^4}$$

Therefore, if the series $\alpha+\beta t+\gamma t^2+etc.$ is expanded into a regular simple series it will result in

$$s=\alpha\left(1-\frac{t^2}{1\cdot 2\cdot 2ab}+\frac{t^4}{1\cdot 2\cdot 3\cdot 4\cdot 4a^2 b^2}-\frac{t^6}{1\cdot 2\cdot 3\cdot\ldots\cdot 6\cdot 8a^3 b^3}+etc.\right)$$

$$+\beta\sqrt{2ab}\left(\frac{t}{\sqrt{2ab}}-\frac{t^3}{1\cdot 2\cdot 3\cdot 2ab\sqrt{2ab}}+\frac{t^5}{1\cdot 2\cdot 3\cdot\ldots\cdot 5\cdot 4a^2 b^2\sqrt{2ab}}-etc.\right)$$

$$+\frac{ab\sqrt{2ab}}{g(a-2b)}\left(\frac{t}{\sqrt{2ab}}-\frac{t^3}{1\cdot 2\cdot 3\cdot 2ab\sqrt{2ab}}+\frac{t^5}{1\cdot 2\cdot 3\cdot 4\cdot 5\cdot 4a^2 b^2\sqrt{2ab}}-etc.\right)$$

$$-\frac{a^2 b}{g(a-2b)}\left(\frac{t}{a}-\frac{t^3}{1\cdot 2\cdot 3\cdot a^3}+\frac{t^5}{1\cdot 2\cdot 3\cdot 4\cdot 5\cdot a^5}-etc.\right)$$

since this series individual parts are summable, the location $s$ will be obtained from a sequence of finite values as $s=\alpha\cos\frac{t}{\sqrt{2ab}}+\beta\sqrt{2ab}\sin\frac{t}{\sqrt{2ab}}+\frac{ab\sqrt{2ab}}{g(a-2b)}\sin\frac{t}{\sqrt{2ab}}-\frac{a^2 b}{g(a-2b)}\sin\frac{t}{a}$ , in this equation, if the constants $\alpha$ and $\beta$ are transformed accordingly, it almost coincide with the one obtained from integration in §27 above.

§32. Let us keep the equation obtained above $s=D\cos\frac{t}{\sqrt{2ab}}+C\sin\frac{t}{\sqrt{2ab}}-\frac{a^2 b}{g(a-2b)}\sin\frac{t}{a}$ and $\sqrt{v}=\frac{D}{\sqrt{2b}}\sin\frac{t}{\sqrt{2ab}}-\frac{C}{\sqrt{2b}}\cos\frac{t}{\sqrt{2ab}}+\frac{ab\sqrt{a}}{g(a-2b)}\cos\frac{t}{a}$, since both of the special cases handled above contain the essentials. Now, let us set that at the beginning, $t=0$, the body rests at $C$, and then, $s=0$, and $\sqrt{v}=0$. Therefore $D=0$; and $C=\frac{ab\sqrt{2ab}}{g(a-2b)}$, and by a direct substitution, results in

$\boldsymbol{s = \frac{ab\sqrt{2ab}}{g(a-2b)}\sin\frac{t}{\sqrt{2ab}} - \frac{a^2b}{g(a-2b)}\sin\frac{t}{a}}$ , and, also, $\boldsymbol{\sqrt{v} = \frac{ab\sqrt{a}}{g(a-2b)}\left(\cos\frac{t}{a} - \cos\frac{t}{\sqrt{2ab}}\right)}$, and from these equations, the position of the body and its velocity at a given time will become known.

§33. To get into the nature of these oscillations more deeply, we will consider various relations between the quantities $a$ and $b$, to which the arcs $\frac{t}{a}$ and $\frac{t}{\sqrt{2ab}}$ render commensurable. First, indeed, let us begin with the maximum value of $b$, and, in this case, the force that depends on the distance $s$ becomes zero. Since in this case, $\sin\frac{t}{\sqrt{2ab}} = \frac{t}{\sqrt{2ab}}$[m] , then, $s = \frac{-at}{2g} + \frac{a^2}{2g}\sin\frac{t}{a}$; and also, $\sqrt{v} = \frac{a\sqrt{a}}{2g}\left(1 - \cos\frac{t}{a}\right)$.[n] Hence,

| If time $\frac{t}{\sqrt{a}}$ | Then, the distance $s$ | And the Velocity $\sqrt{v}$ |
|---|---|---|
| $\frac{0c}{\sqrt{a}}$ | $0$ | $0$ |
| $\frac{c}{\sqrt{a}}$ | $\frac{-ac+a^2}{2g}$ | $\frac{a\sqrt{a}}{2g}$ |
| $\frac{2c}{\sqrt{a}}$ | $\frac{-2ac}{2g}$ | $\frac{2a\sqrt{a}}{2g}$ |
| $\frac{3c}{\sqrt{a}}$ | $\frac{-3ac-a^2}{2g}$ | $\frac{a\sqrt{a}}{2g}$ |
| $\frac{4c}{\sqrt{a}}$ | $\frac{-4ac}{2g}$ | $0$ |
| $\frac{5c}{\sqrt{a}}$ | $\frac{-5ac-a^2}{2g}$ | $\frac{a\sqrt{a}}{2g}$ |

§34. Therefore, for this case, in which we consider $b$ infinite, the body will continuously advance from $C$ towards the right beyond $CB$, in an alternate accelerated and retarded motion. Although, in this case, the oscillations do not occur, yet, it seems appropriate to start the analysis from it, so that the link between the motions originated in this manner becomes clearer, as long as $b$ decreases gradually to a smaller value. Let us set $b = \frac{n^2a}{2}$, such that $\sqrt{2ab} = na$; whence, $s = \frac{n^2a^2}{2g(n^2-1)}\left(\sin\frac{t}{a} - n\sin\frac{t}{na}\right)$, and also, $\sqrt{v} = \frac{n^2a\sqrt{a}}{2g(n^2-1)}\left(\cos\frac{t}{na} - \cos\frac{t}{a}\right)$: in these expressions, the sine and cosine of the arcs $\frac{t}{a}$ and $\frac{t}{na}$ can be interchanged, whenever $n$ is a rational number.

§35. Let us continuously decrease the value of $n$, from the previous value, which was considered infinite, until it approaches the case for $n = 1$; reducing the equation to this particular form $s = \frac{-a^2}{4g}\sin\frac{t}{a} +$

---

[m] Since $b$ is being considered a very large quantity, the arc $\frac{t}{\sqrt{2ab}}$ would be very small, and, in this case, the sine of a very small arc can be taken as the arc itself. Therefore, $s = \frac{ab\sqrt{2ab}}{g(a-2b)}\sin\frac{t}{\sqrt{2ab}} - \frac{a^2b}{g(a-2b)}\sin\frac{t}{a} = \frac{ab}{g(a-2b)} - \frac{a^2b}{g(a-2b)}\sin\frac{t}{a}$. Since $\frac{ab}{g(a-2b)} \approx \frac{-a}{2g}$ and $-\frac{a^2b}{g(a-2b)} \approx \frac{a^2}{2g}$, then, finally, $s = \frac{-at}{2g} + \frac{a^2}{2g}\sin\frac{t}{a}$.

[n] Here, the symbology of the expression that appears in the original manuscript could not be identified, and the logical expression was written instead.

$\frac{at}{4g}\cos\frac{t}{a}$, and also, $\sqrt{v} = \frac{t\sqrt{a}}{4g}\sin\frac{t}{a}$ ;[o] therefore, in this case, the oscillations eventually last indefinitely: and contrary [to the previous case], the motion will endure.

| If time $\frac{t}{\sqrt{a}}$ | Then, the distance s | And the Velocity $\sqrt{v}$ |
|---|---|---|
| $\frac{0c}{\sqrt{a}}$ | $0$ | $0$ |
| $\frac{c}{\sqrt{a}}$ | $-\frac{a^2}{4g}$ | $+\frac{c\sqrt{a}}{4g}$ |
| $\frac{2c}{\sqrt{a}}$ | $-\frac{2ac}{4g}$ | $0$ |
| $\frac{3c}{\sqrt{a}}$ | $+\frac{a^2}{4g}$ | $-\frac{3c\sqrt{a}}{4g}$ |
| $\frac{4c}{\sqrt{a}}$ | $+\frac{4ac}{4g}$ | $0$ |
| $\frac{5c}{\sqrt{a}}$ | $-\frac{a^2}{4g}$ | $+\frac{5c\sqrt{a}}{4g}$ |

§36. Having the nearly extreme cases been unfolded, namely, $n = \infty$ $and$ $n = 1$, let us see how much the intermediate cases, for which we put successive integers for $n$ which differ from the extremes. Let us consider $n = 2$, or $b = 2a$; then, $s = \frac{2a^2}{3g}\left(\sin\frac{t}{a} - 2\sin\frac{t}{2a}\right)$, and also, $\sqrt{v} = \frac{2a\sqrt{a}}{3g}\left(\cos\frac{t}{2a} - \cos\frac{t}{a}\right)$. Therefore, whenever $t = 4ic$, then, $s = 0$; whereas the velocity will vanish whenever $t = \frac{8ic}{3}$, designating $i$ as any integer number.

| If time $\frac{t}{\sqrt{a}}$ | Then, the distance s | And the Velocity $\sqrt{v}$ |
|---|---|---|
| $and\ t = \frac{0c}{3}$ | $0$ | $0$ |
| $t = \frac{4c}{3}$ | $\frac{-2a^2}{3g}\sin\frac{2c}{3a}$ | $\frac{+4a\sqrt{a}}{3g}\cos\frac{2c}{3a}$ |
| $t = \frac{8c}{3}$ | $\frac{-6a^2}{3g}\sin\frac{2c}{3a}$ | $0$ |
| $t = \frac{12c}{3}$ | $0$ | $\frac{-4a\sqrt{a}}{3g}$ |
| $t = \frac{16c}{3}$ | $\frac{+6a^2}{3g}\sin\frac{2c}{3a}$ | $0$ |
| $t = \frac{20c}{3}$ | $\frac{+2a^2}{3g}\sin\frac{2c}{3a}$ | $\frac{+4a\sqrt{a}}{3g}\cos\frac{2c}{3a}$ |
| $t = \frac{24c}{3}$ | $0$ | $0$ |

§37. Therefore, the motion revolutions recommence after the time $\frac{8c}{\sqrt{a}}$ has elapsed, or after the circumference has been completed twice; meanwhile three oscillations have been completed, the

[o] These two expressions, which have been derived from those presented in §34, have been confirmed by the Translator.

middle of which throughout the distance is twice as large as the remaining[p]. Similarly, if $n = 3$, for which the same periods return after the time $\frac{32c}{\sqrt{a}}$ has elapsed, or after the circumference has been completed thrice: and so on, until $n = \infty$, where there are no periods of revolution, with the body in that location moving perpetually to infinite. For $n = 3$, the velocity $\sqrt{v}$ completely vanishes, whenever $t = 3ic$: and, if $n = 4$, the velocity of the body vanishes, for cases when $t = \frac{16ic}{3}$ and $t = \frac{16ic}{5}$. Then, be set $t = \frac{16ic}{15}$, and if $i$ is substituted for a sequence of integers, the velocity of the body will turn out zero for cases where $i$ equals to: $0, 3, 5, 6, 9, 10, 12, 15, 18, 20, 21, 24, 25,$ etc. they will differ for: $3, 2, 1, 3, 1, 2, 3, 3, 2, 1, 3, 1,$[q] after the time 16c has elapsed, the same period will be repeated, and seven times in one period, the velocity will be null, such that unequal oscillations will be contained in just one period: and if, in fact, one more oscillation is added between two limits, the speed with which is $= 0$.

§38. These oscillations will become more regular if $n < 1$ and $\frac{t}{n}$ is an integer. Let us set $b = \frac{a}{2n^2}$,[r] such that $\sqrt{2ab} = \frac{a}{n}$, and then $\boldsymbol{s = \frac{a^2}{2g(n^2-1)}\left(\frac{1}{n}\sin\frac{nt}{a} - \sin\frac{t}{a}\right)}$ , and, also, $\boldsymbol{\sqrt{v} = \frac{a\sqrt{a}}{2g(n^2-1)}\left(\cos\frac{t}{a} - \cos\frac{nt}{a}\right)}$. Therefore, the velocity of the body will completely vanish whenever $t = \frac{4ic}{n \mp 1}$. Moreover, the body will not return to the point $C$, for which $s = 0$, unless $t = 2ic$. But, if it is assumed that $t = \frac{4ic}{n-1}$, then, $\boldsymbol{s = \frac{-a^2}{2gn(n+1)}\sin\frac{t}{a}}$, however, if $t = \frac{4ic}{n+1}$, then, $\boldsymbol{s = \frac{-a^2}{2gn(n-1)}\sin\frac{t}{a}}$ . Thus, the results in following tables were obtained by putting in these formulas successive values for $n$ such as $2, 3, 4, 5, etc.$, which would allow the oscillatory motion of the body driven by two forces to be known.

$$Set\ n = 2\ or\ b = \frac{a}{8}$$

| If time $\frac{t}{\sqrt{a}}$ | Then, the distance s | And the Velocity $\sqrt{v}$ |
|---|---|---|
| $set\ t = 0c$ | 0 | 0 |
| $set\ t = c$ | $\frac{-a^2}{6g}$ | $\frac{a\sqrt{a}}{6g}$ |
| $set\ t = \frac{4}{3}c$ | $-\frac{a^2}{4g}\sin\frac{2c}{3a}$ | 0 |
| $set\ t = 2c$ | 0 | 98 |
| $set\ t = \frac{8}{3}c$ | $+\frac{a^2}{4g}\sin\frac{2c}{3a}$ | 0 |
| $set\ t = 3c$ | $+\frac{a^2}{6g}$ | $+\frac{a\sqrt{a}}{6g}$ |
| $set\ t = 4c$ | 0 | 0 |

[p] Not clear what is meant here.

[q] It is not clear what this sequence of numbers refers to.

[r] The values that appear in the next tables were derived under this condition.

$$Set\ n = 3\ or\ b = \frac{a}{18}$$

| If time $\frac{t}{\sqrt{a}}$ | Then, the distance s | And the Velocity $\sqrt{v}$ |
|---|---|---|
| $set\ t = 0c$ | 0 | 0 |
| $set\ t = c$ | $-\frac{a^2}{12g}$ | 0 |
| $set\ t = 2c$ | 0 | 0 |
| $set\ t = 3c$ | $+\frac{a^2}{12g}$ | 0 |
| $set\ t = 4c$ | 0 | 0 |

$$Set\ n = 4\ or\ b = \frac{a}{32}$$

| If time $\frac{t}{\sqrt{a}}$ | Then, the distance s | And the Velocity $\sqrt{v}$ |
|---|---|---|
| $set\ t = 0$ | 0 | 0 |
| $set\ t = \frac{4}{5}c$ | $-\frac{a^2}{24g}\sin\frac{4c}{5a}$ | 0 |
| $set\ t = c$ | $-\frac{a^2}{40g}$[s] | $-\frac{a\sqrt{a}}{30g}$ |
| $set\ t = \frac{4}{3}c$ | $-\frac{a^2}{40g}\sin\frac{2c}{3a}$ | 0 |
| $set\ t = \frac{8}{5}c$ | $\frac{-a^2}{24g}\sin\frac{2c}{5a}$ | 0 |
| $set\ t = 2c$ | 0 | $-\frac{a\sqrt{a}}{15g}$ |
| $set\ t = \frac{12}{5}c$ | $+\frac{a^2}{24g}\sin\frac{2c}{5a}$ | 0 |
| $set\ t = \frac{8}{3}c$ | $+\frac{a^2}{40g}\sin\frac{2c}{3a}$ | 0 |
| $set\ t = 3c$ | $+\frac{a^2}{40g}$[t] | $-\frac{a\sqrt{a}}{30g}$ |
| $set\ t = \frac{16}{5}c$ | $+\frac{a^2}{24g}\sin\frac{4c}{5a}$ | 0 |
| $set\ t = 4c$ | 0 | 0 |

$$Set\ n = 5\ or\ b = \frac{a}{50}$$

[s] Incorrectly written as $-\frac{a^2}{30g}$

| If time $\frac{t}{\sqrt{a}}$ | Then, the distance s | And the Velocity $\sqrt{v}$ |
|---|---|---|
| $set\ t = 0$ | $0$ | $0$ |
| $set\ t = \frac{2}{3}c$ | $-\frac{a^2}{40g}\sin\frac{2c}{3a}$ | $0$ |
| $set\ t = c$ | $-\frac{a^2}{60g}$ | $0$ |
| $set\ t = \frac{4}{3}c$ | $-\frac{a^2}{40g}\sin\frac{2c}{3a}$ | $0$ |
| $set\ t = 2c$ | $0$ | $0$ |
| $set\ t = \frac{12}{5}c$ | $+\frac{a^2}{24g}\sin\frac{2c}{5a}$ | $0$ |
| $set\ t = \frac{8}{3}c$ | $+\frac{a^2}{40g}\sin\frac{2c}{3a}$ | $0$ |
| $set\ t = 3c$ | $+\frac{a^2}{60g}$ | $0$ |
| $set\ t = \frac{10}{3}c$ | $+\frac{a^2}{40g}\sin\frac{2c}{3a}$ | $0$ |
| $set\ t = 4c$ | $0$ | $0$ |

§39. Among all these cases, the one which deserves particular attention is that for which $2b = a$:[u] in which the oscillation distance eventually grows up to infinite: this effect is most remarkable, since it is generated by finite forces. Therefore, if it would be possible to revoke, as usual, a perpetual mobile, this case can be considered as a potential plan of a derivation: indeed, it is now compared with a pendulum oscillating in a cycloid, as the impulses originated by gravity versus the point of equilibrium are related to the travelled distances. Hence, if such a pendulum is applied to an automaton, which produces another time-dependent force, the force of the oscillations could increase to such a degree that the portion renovating the stretching of the automaton could surpass, whenever necessary, that expended to resistance and friction, such that if the oscillations do not increase, yet, the given quantities are constantly conserved.

§40. If we now ask for the cause of why this is the only case where the oscillations increase continuously, we would find no other reason than the case where the time of a unique complete oscillation is composed of one advance and one retardation, which is produced by the action of a unique force which depends on the distance, where [the time of a complete oscillation] is expressed by the whole circumference $FDHE$. If the body is only driven by the force $\frac{s}{b}$, then the time of one complete oscillation from one advance and one retardation will be constant and $= 2\pi\sqrt{2b} = \frac{4c}{a}\sqrt{2b}$, because

[t] Incorrectly written as $+\frac{a^2}{30g}$

[u] This is the case dealt with in §35.

$1:\pi = a:2c$. However, the time expent along the whole circumference is expressed as $= \frac{4c}{\sqrt{a}}$; then, for equal times, it is necessary that $2b = a$, which is the case just considered.

§41. Hence, the nature of the difference in behaviors that we have observed among the oscillations in the last cases can now be analyzed in more detail. In fact, this difference depends in part on the value of the letter $g$, since, indeed, there are no other ways for a diversity of oscillations to be induced, unless that through it, greater distances are accomplished the less the value of $g$, and, indeed, the computation of the motion and of time remains similar. On the other hand, the most important part in changing the nature of the oscillations, is the different values for the letters $a$ and $b$, through which the time of oscillations generated individually by both forces are defined. Indeed, the time of a unique oscillation due to just the force $\frac{s}{b}$, to the time of a unique oscillation due to just the force $\frac{y}{g}$ is as $\sqrt{2b}$ to $\sqrt{a}$. From which, it becomes clear that the more this ratio is reduced, the more the resulting oscillations become irregular.